\documentclass[preprint]{aastex631}

\setcitestyle{notesep={ }}

\begin{document}

\title{The Impacts of Magnetogram Projection Effects on Solar Flare Forecasting}

\author[0000-0003-3493-9174]{Griffin T. Goodwin}

\author[0000-0002-4001-1295]{Viacheslav M. Sadykov}

\author[0000-0001-8078-6856]{Petrus C. Martens}
\affiliation{Physics \& Astronomy Department, Georgia State University, Atlanta, GA 30303, USA; \href{mailto:ggoodwin5@gsu.edu}{ggoodwin5@gsu.edu}}


\begin{abstract}
 This work explores the impacts of magnetogram projection effects on machine learning-based solar flare forecasting models. Utilizing a methodology proposed by \cite{falconer2016new}, we correct for projection effects present in Georgia State University’s Space Weather Analytics for Solar Flares (SWAN-SF) benchmark data set. We then train and run a support vector machine classifier on the corrected and uncorrected data, comparing differences in performance. Additionally, we provide insight into several other methodologies that mitigate projection effects, such as stacking ensemble classifiers and active region location-informed models. Our analysis shows that data corrections slightly increase both the true positive (correctly predicted flaring samples) and false positive (non-flaring samples predicted as flaring) prediction rates, averaging a few percent. Similarly, changes in performance metrics are minimal for the stacking ensemble and location-based model. This suggests that a more complicated correction methodology may be needed to see improvements. It may also indicate inherent limitations when using magnetogram data for flare forecasting.
\end{abstract}

\keywords{Space weather (2037), Solar flares (1496), Support vector machine (1936)}


\section{Introduction} \label{sec:intro}
Active region (AR) vector magnetograms are frequently used to forecast solar flares as they provide insight into one of the only reservoirs capable of driving the flaring process: free magnetic energy. By deriving a set of physics-based parameters from the data, we can produce a comprehensive set of preflare features that can be easily integrated into machine learning (ML) models. However, as ARs traverse farther from the solar disk center, the quality of magnetic field data degrades substantially. Near the solar limb, AR vector magnetograms and their properties (magnetic flux, current helicity, etc) suffer from systematic observational trends unrelated to the evolution of the AR itself, known as projection effects. These trends appear due to line height formation effects, foreshortening of magnetogram pixels, and increased noise which may make an AR appear stronger or weaker than it truly is \citep{Kitiashvili_Couvidat_Lagg_2015,falconer2016new}. This complicates forecasting, especially for events near the classification border, potentially introducing unnecessary false positive and negative predictions. Correcting for these effects appears to be imperative in ensuring a well-performing model, yet, as far as we know, no studies have explored multi-parameter corrections thoroughly. In this work, our goal is to employ a methodology similar to that of \cite{falconer2016new}, which was originally used for unsigned magnetic flux corrections, to remove the projection effects for the AR features present in Georgia State University’s Space Weather Analytics for Solar Flares benchmark data set \citep{SWANSF_2020,Angryk_2020}, and examine the impacts these corrections may have on forecasting performance. Additionally, we make use of two other correction methodologies, a stacking ensemble classifier, and an active region location-informed model, to reduce the influence of projection effects. Interestingly, as far as we are aware, these techniques have not been explored much in the literature. We would like to note that ideally projection effect corrections should be applied directly to the data source itself (the magnetograms), however, due to the significant amount of effort required for computing the derived magnetogram features, this methodology is beneficial to consider.


The remaining sections of this work are organized as follows: Section \ref{sec:data} highlights the data used in our predictions. Section \ref{sec:method} describes the methodology used to correct the projection effects. Finally, Sections \ref{sec:results} and \ref{sec:summary} detail our results and conclusions from this work.


\section{Data} \label{sec:data}
\begin{figure}[htb]
    \centering
    \includegraphics[width=.85\textwidth]{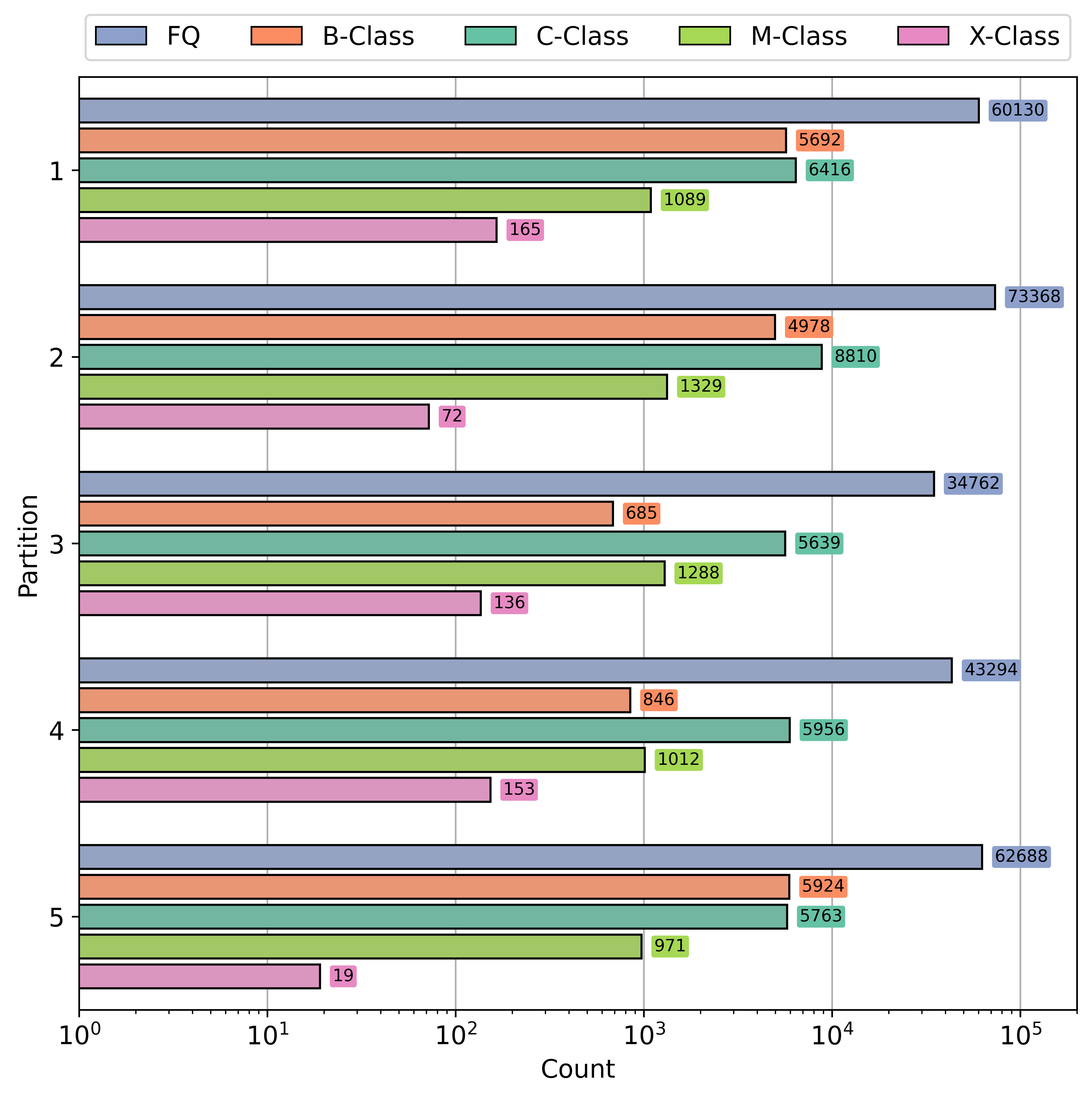}
    \caption{The flare class distribution for each partition within the Space Weather Analytics for Solar Flares benchmark data set. Bars are color-coded by the flaring class: Flare quiet (blue), B- (orange), C- (teal), M- (green), and X-class (pink).}
    \label{fig:classDistribution}
\end{figure}

Georgia State's Space Weather Analytics for Solar Flares (SWAN-SF) benchmark data set \citep{SWANSF_2020, Angryk_2020} is an ML-ready compilation of Solar Dynamics Observatory Helioseismic and Magnetic Imager \citep[SDO HMI,][]{scherrer2012helioseismic} AR vector magnetogram parameter time series spanning part of Solar Cycle 24 (May 2010 - August 2018). Each file contains 12 hours' worth of data at a 12-minute cadence, with twenty-four physics-based parameters calculated at each timestamp (see Table \ref{tab:magparams}). Every file is labeled based on the strongest flaring instance to occur in the following 24 hours, with M- and X-class flares deemed as positive (or flaring) events, while weaker B-, C-class, and completely flare quiet instances are considered negative (or non-flaring) events. SWAN-SF is divided into five temporally segmented partitions, each containing approximately the same number of M- and X-class flares \citep{Ahmadzadeh_Aydin_Georgoulis_Kempton_Mahajan_Angryk_2021}. This allows for quick and efficient training/testing of ML classifiers, which we take advantage of in this work. Figure \ref{fig:classDistribution} depicts the distribution of flaring/non-flaring classes within each partition. If the reader would like to learn more about how SWAN-SF was constructed, please see \cite{Angryk_2020}. 
\begin{deluxetable}{|c|l|c|}
\tabletypesize{\scriptsize}
\def\arraystretch{1.6}
\tablehead{\colhead{\textbf{Parameter}} & \colhead{\textbf{Description}} & \colhead{\textbf{Formula}}}
\startdata
\texttt{ABSNJZH}                                                  & Absolute value of the net current helicity in G$^2$/m & $H_{c_{a b s}} \propto\left|\sum B_z \cdot J_z\right|$\\ \hline
\texttt{EPSX}    & Sum of X-component of normalized Lorentz force & $\delta F_x \propto \frac{\sum B_xB_z}{\sum B^2}$   \\ \hline
\texttt{EPSY}    & Sum of Y-component of normalized Lorentz force & $\delta F_y \propto \frac{-\sum B_yB_z}{\sum B^2}$  \\ \hline
\texttt{EPSZ}    & Sum of Z-component of normalized Lorentz force & $\delta F_z \propto \frac{\sum (B_x^2+B_y^2-B_z^2)}{\sum B^2}$  \\ \hline
\texttt{MEANALP} & Mean twist parameter                           & $\alpha_{total} \propto \frac{\sum J_z \cdot B_z}{\sum B_z^2}$  \\ \hline
\texttt{MEANGAM} & Mean inclination angle                         & $\overline{\gamma} = \frac{1}{N}\sum arctan(\frac{B_h}{B_z})$  \\ \hline
\texttt{MEANGBH} & Mean value of the horizontal field gradient    & $\overline{\nabla B_h} = \frac{1}{N}\sum \sqrt{(\frac{\partial B_h}{\partial x}+\frac{\partial B_h}{\partial y})}$  \\ \hline
\texttt{MEANGBT} & Mean value of the total field gradient         & $\overline{|\nabla B_{tot}|} = \frac{1}{N}\sum \sqrt{(\frac{\partial B}{\partial x}+\frac{\partial B}{\partial y})}$  \\ \hline
\texttt{MEANGBZ} & Mean value of the vertical field gradient      & $\overline{\nabla B_z} = \frac{1}{N}\sum \sqrt{(\frac{\partial B_z}{\partial x}+\frac{\partial B_z}{\partial y})}$   \\ \hline
\texttt{MEANJZD} & Mean vertical current density                  & $\overline{J_z} \propto \frac{1}{N}\sum (\frac{\partial B_y}{\partial x}-\frac{\partial B_x}{\partial y})$ \\ \hline
\texttt{MEANJZH} & Mean current helicity                          & $\overline{H_c} \propto \frac{1}{N}\sum B_z \cdot J_z$ \\ \hline
\texttt{MEANPOT} & Mean photospheric excess magnetic energy density & $\overline{\rho} \propto \frac{1}{N}\sum (B^{Obs}-B^{Pot})^2$ \\ \hline
\texttt{MEANSHR} & Mean shear angle                                & $\overline{\Gamma} = \frac{1}{N} \sum arccos \left( \frac{B^{Obs}\cdot B^{Pot}}{|B^{Obs}||B^{Pot}|} \right)$\\ \hline
\texttt{R\_VALUE}                                               & Total unsigned flux around high gradient polarity inversion lines using the B$_{los}$ component & $\Phi = \sum |B_{los}|\cdot dA \text{ (within R mask)}$ \\ \hline
\texttt{SAVNCPP}                                                  & Sum of the absolute value of the net current per polarity                                 & $J_{z_{sum}}  \propto |\sum^{B_z^+}J_zdA| + |\sum^{B_z^-}J_zdA|$     \\ \hline
\texttt{SHRGT45} & Area with shear angle greater than 45 degrees  & $\frac{\text{Area with Shear } > 45^\circ}{\text{Total Area}}$  \\ \hline
\texttt{TOTBSQ}  & Total magnitude of Lorentz force               & $F \propto \sum B^2$  \\ \hline
\texttt{TOTFX}  & Sum of X-component of Lorentz force             & $F_x \propto \sum B_x B_z dA$ \\ \hline
\texttt{TOTFY}   & Sum of Y-component of Lorentz force            & $F_y \propto \sum B_y B_z dA$  \\ \hline
\texttt{TOTFZ}   & Sum of Z-component of Lorentz force            & $F_z \propto \sum (B_x^2+B_y^2-B_z^2) dA$  \\ \hline
\texttt{TOTPOT}  & Total photospheric magnetic energy density     & $\rho_{tot} \propto \sum(\overrightarrow{B}^{Obs} - \overrightarrow{B}^{Pot})^2 dA$ \\ \hline
\texttt{TOTUSJH} & Total unsigned current helicity                & $H_{c_{total}} \propto \sum B_z \cdot J_z$  \\ \hline
\texttt{TOTUSJZ} & Total unsigned vertical current                & $J_{z_{total}} = \sum |J_z|dA$  \\ \hline
\texttt{USFLUX}  & Total unsigned flux in Maxwells                & $\Phi = \sum|B_z|dA$  \\ \hline
\enddata
\caption{The derived magnetic field parameters in the Space Weather Analytics for Solar Flares benchmark data set. These parameters have been curated by \cite{bobra2015solar}. This table has been adapted from \cite{Angryk_2020}.}
\label{tab:magparams}
\end{deluxetable}

We want to make clear that our primary focus in this work is to study the impacts of projection effects on flare forecasting, rather than produce the best-performing classifier. Thus, to simplify and accelerate our training/testing process, we employ a feature-vector-based (contrary to a time series-based) forecaster, which requires us to reduce each multivariate time series file to a single data point. To achieve this, we calculate the descriptive statistics for each of the twenty-four magnetogram parameters in a given 12 hr file. This includes the mean, median, standard deviation, variance, maximum, minimum, skewness, and kurtosis. Then, to capture aspects pertinent to the evolution of the time series, we record the last value in the file, the mean of the gradient derivative, and the standard deviation of the gradient derivative. For files with missing data, we linearly interpolated beforehand. This method saves us significant time, reducing each file from 60 timestamps by 24 parameters (or 1440 dimensions) to 1 timestamp by 264 parameters, i.e., 11 parameters for each of the 24 magnetogram parameters.

\section{Methodology} \label{sec:method}
 In this work, we follow a similar methodology introduced by \cite{falconer2016new} to correct for projection effects. By normalizing the magnetic field data for a given AR to the value near the central meridian (where projection effects are minimal), combining the results across many different ARs, and employing a polynomial fit, we create a curve that describes the severity of the projection effects as a function of heliocentric angle ($\theta$, the angle between an AR's line-of-sight and surface normal). For any new AR, we can then remove the projection effects by dividing the data by the polynomial value at the AR’s current heliocentric angle, rescaling the data back to unity (or the estimated unprojected value). In the following sections, we will provide details on how we selected our ``benchmark" ARs for modeling the polynomials (see Section \ref{sec:ARSelect}), how we generated the polynomial fits for each correction (see Section \ref{sec:fittingPoly}), how we trained our ML models (see Section \ref{sec:MLModels}), and the performance metrics we used to analyze our results (see Section \ref{sec:performanceMetrics}).

\subsection{Active Region Selection} \label{sec:ARSelect}
\begin{figure}[htb]
    \centering
    \includegraphics[width=\textwidth]{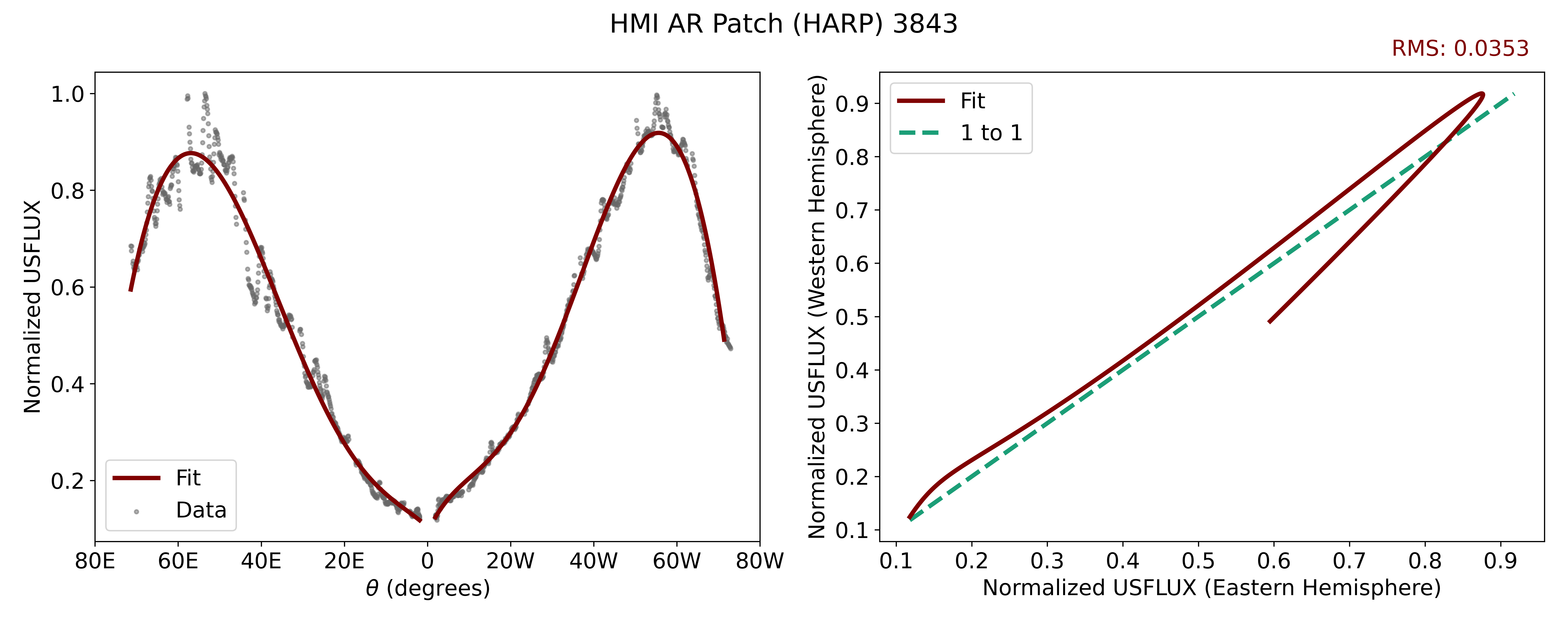}
    \caption{(\textit{Left}) The normalized unsigned magnetic flux for HARP 3843 versus its heliocentric angle ($\theta$) in degrees. The data in grey are normalized to the max unsigned magnetic flux. A spline is fit through the data in the eastern hemisphere (left of 0$^\circ$) and the western hemisphere (right of 0$^\circ$) in maroon. Notice the highly symmetric nature of the data across the traversal. This indicates that the flux variations we observe are mostly dominated by projection effects. (\textit{Right}) In maroon, the unsigned magnetic flux spline from the western hemisphere is plotted versus the unsigned magnetic flux spline from the eastern hemisphere (at the same heliocentric angle). A one-to-one line is shown in green. The root mean squared value between the fit data and one-to-one line is shown in the top right.}
    \label{fig:ARSelection}
\end{figure}
 The first step in the correction process is to identify the ``benchmark" ARs that will be used to model the average projection effects across the solar disk. Ideally, these ARs should demonstrate minimal variations in their strength over their lifetimes. We can ensure this by analyzing an AR's unsigned magnetic flux profiles, only selecting those that show a highly symmetric evolution across the solar meridian. This guarantees that any observed growths/decays in the data are (mostly) due to the projection effects themselves, rather than any intrinsic evolution of the AR that may occur. We invoke the following methodology to quantitatively analyze the symmetry of the flux profile for a given AR. First, we down-select ARs from the 4098 present in SWAN-SF based on the criterion that their data spans most of the solar disk. Since we will only be making forecasts for ARs with $\theta$s out to 70$^\circ$, we select ARs that have data within longitudes and $\theta$s of 73$^\circ$ (a 3$^\circ$ buffer is included to ensure a good final polynomial fit). We also want the ARs to pass relatively close to the disk center, so we include an additional criterion that data must exist within $\theta$s of 30$^\circ$. From here, we take each of the 398 down-sampled ARs and fit two (one for each hemisphere) 5th-degree splines through the unsigned magnetic flux normalized to the maximum value along the traversal, as a function of the $\theta$. We then take the data from the fit of the western hemisphere and plot it against the data from the eastern hemisphere (at the same $\theta$). We also include a one-to-one line that depicts a perfectly symmetric evolution (see Figure \ref{fig:ARSelection}). By calculating the root mean squared (RMS) error between the fit data and the one-to-one line, we can quantify how symmetric the flux evolution is across hemispheres. A low RMS should correspond to a good one-to-one fit. Through visual inspection, we settled on an RMS cutoff of $<$0.3 for our final sample of ARs. This ensures that we have a sufficient number of ARs to work with while removing the egregiously asymmetric ARs. In the end, we were left with 260 ARs.

\subsection{Fitting and Applying the Corrective Polynomials} \label{sec:fittingPoly}

Now that we have selected our ``benchmark" ARs, we can begin generating the correction polynomials for each magnetic field parameter in Table \ref{tab:magparams}. Our approach consists of two stages. First, we establish a ``prefit" polynomial that will calibrate the final fit. This ``prefit" curve is constructed using only ARs that pass within 10$^\circ$ of the solar meridian. For each of these ARs, we normalize the data based on the mean value of the observed magnetic field parameter between the AR's minimum $\theta$ and the minimum plus 3$^\circ$. This ensures that near 0$^\circ$, where we expect minimal projection effects, the correction is set to unity. Then, the rescaled data beyond this $\theta$ range models the strength of the projection effects. Combining all of our ``prefit" normalized AR data, we can apply a bootstrapping technique to generate a fit with uncertainties. This method involves resampling our ``prefit" AR data 500 times, with replacement. More specifically, say we had $x$ number of ARs within our ``prefit" data. Within a single bootstrapped sample, $x$ random AR selections are made from our original pool with the possibility of selecting the same region multiple times. We repeat this process 500 times and apply a 7th-degree polynomial fit through the data of each of our bootstrapped samples. This fit is defined as follows:
\begin{equation}
    y = 1+\beta_0\theta+\beta_1\theta^2+\beta_2\theta^3+\beta_4\theta^5+\beta_5\theta^6+\beta_6\theta^7
\end{equation}
where $\beta_n$ are constants, $\theta$ is the observed heliocentric angle, and $y$ is the correction factor. The average bootstrapped fit is then used as the final correction polynomial for the ``prefit" data. The reason we create ``prefit" polynomials first is that ARs with minimum $\theta$s greater than 10$^\circ$ should not necessarily be normalized to unity. In principle, these ARs will already have some degradation due to projection effects. Thus, we normalize their data to the average ``prefit" polynomial value between the minimum $\theta$ and the minimum plus 3$^\circ$. Once this is done, we re-run the bootstrapping method, averaging the curves to achieve our final fit for each magnetic field parameter. Having these polynomials then allows us to generate our corrected SWAN-SF data set. By iterating through each time series file, rescaling the 24 magnetogram parameters by their corrective factors at the appropriate $\theta$, and recalculating the descriptive statistics, we should largely remove any projection effects within the data.

We want to note that because some parameters are signed (\texttt{TOTFX/Y/Z}, \texttt{ESPX/Y/Z}, \texttt{MEANJZH}, \texttt{MEANJZD}, \texttt{MEANALP}), we took the absolute value of these features when generating our polynomial fits. The absolute value was not used during forecasting. Additionally, the \texttt{R\_VALUE} feature is dependent on the polarity inversion line (PIL) of the AR. Since not all ARs have a distinct PIL, some time series may contain \texttt{R\_VALUE} = 0, which may interfere with our fit. Thus, data points where \texttt{R\_VALUE} = 0 were removed during fitting.  

\begin{figure}[htb]
    \centering
    \includegraphics[width=\textwidth]{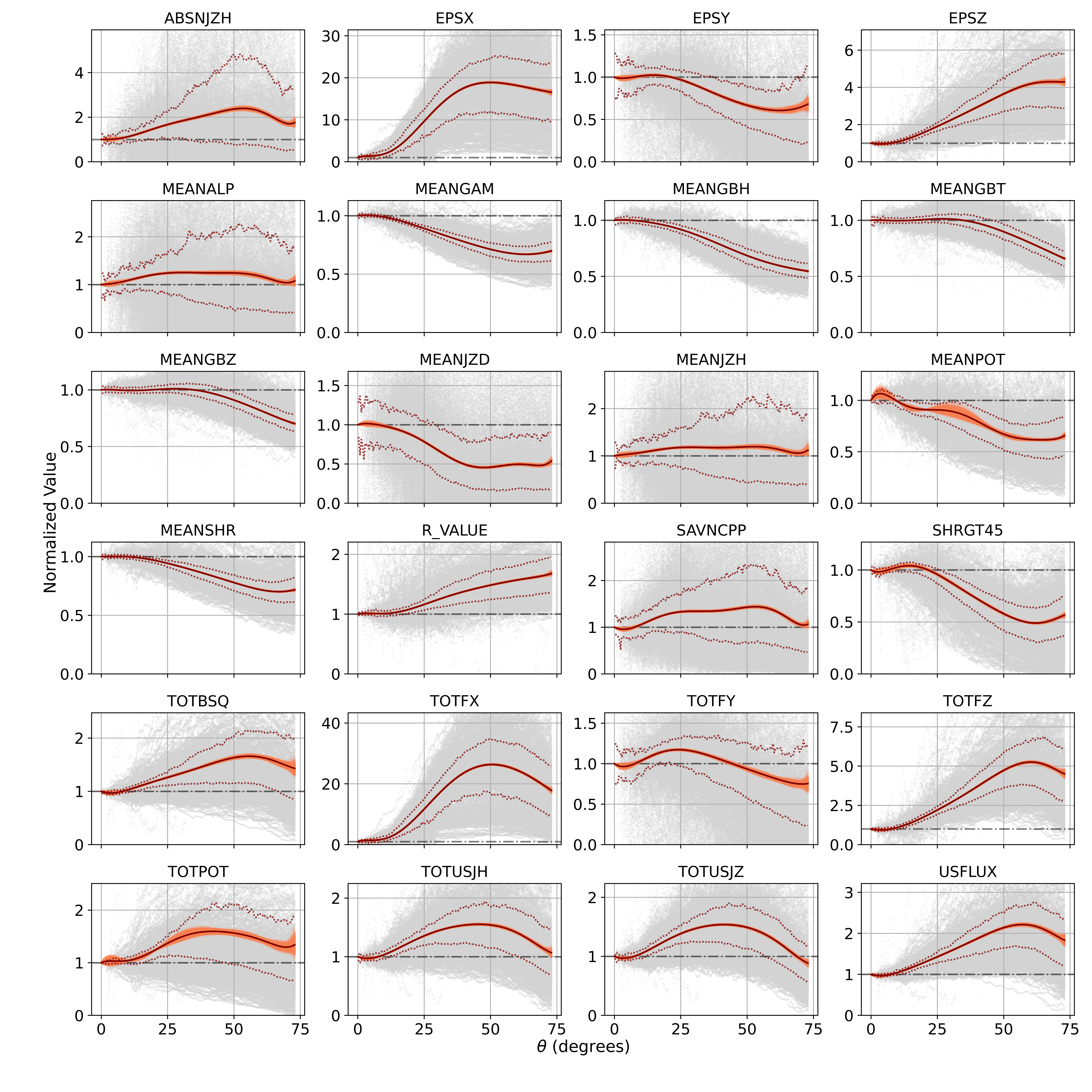}
    \caption{The correction polynomials for each of the 24 derived magnetic field parameters measured in normalized units as a function of heliocentric angle ($\theta$). The solid maroon line represents the average fit from the bootstrapped fits in orange for the data in grey. The dashed maroon lines highlight the median absolute deviation of the data above and below the polynomial fit. The black dashed-dotted line emphasizes the point where the correction factor is 1. Please note that often the bootstrapped fits (orange) are relatively similar to the average fit (maroon), thus the curves can occasionally be hard to identify (e.g. \texttt{MEANSHR}).}
    \label{fig:projectioneffects}
\end{figure}

Figure \ref{fig:projectioneffects} highlights the results of the fitting process, while Table \ref{tab:constants} provides the constants for each fit. The grey dots show the normalized data for the 260 ``benchmark" ARs, the solid maroon line is the average fit of the bootstrapped fits in orange, the dotted maroon line shows the median absolute deviation from the fit, and the black dashed line emphasizes where the correction factor is 1. From the figure, we see that each parameter is influenced differently by projection effects, with correction factors ranging from as low as 0.5 (\texttt{MEANJZD}) up to 25 (\texttt{TOTFX}) at certain $\theta$s. 

The shape of the polynomial tends to land within two categories, either a gradual fall-off at larger $\theta$s or a peak occurring between 30$^\circ$ and 70$^\circ$. Of course, there are a few oddball parameters such as \texttt{MEANALP}, \texttt{MEANJZH}, \texttt{R\_VALUE}, and \texttt{TOTFY} that do not fall into either of these categories. Those that are associated with the hump are directly related to the radial field component of the magnetic field as well as the masking area ($dA$) of the AR (eg. \texttt{USFLUX}), while those with a gradual drop-off tend to be directly related to the horizontal component of the magnetic field (eg. \texttt{MEANGBH)}. As ARs move towards the limb, the radial field component ($B_z$ in the local Cartesian coordinate system rotating with the AR) increasingly influences the transversal field measurements (perpendicular to the line of sight), while the horizontal field components ($B_x$,$B_y$) begin to dominate the line of sight measurements. In the HMI magnetograms, the line of sight component of the field has an error of 5-10 G, while the transverse field noise is on the order of 100 G \citep{liu2012comparison,hoeksema2014helioseismic,falconer2016new}. Consequently, this means near the limb, the increased noise and contribution of the transversal component will expand the ARs masking region. This synthetically increases the features correlated to $B_z$ and $dA$. For the parameters dependent on the horizontal field components, the drop-off likely occurs due to a combination of foreshortening and an increased contribution of the line of sight observation (which has less noise) when making horizontal field measurements. The reduced noise near the limb should decrease the average measurements of the horizontal component, compared to those made near the central meridian, where noise is high.  

Finally, when analyzing the goodness of fit, we find that it varies widely depending on the parameter. For example, \texttt{MEANGAM}, \texttt{MEANBH}, \texttt{MEANGBT}, \texttt{MEANGBZ}, and \texttt{MEANSHR} all have relatively clear trends, while parameters such as \texttt{ABSNJZH}, \texttt{EPSY}, \texttt{MEANALP}, \texttt{MEANJZD}, \texttt{MEANJZH}, \texttt{SAVNCPP}, and \texttt{TOTFY} show no obvious fit and are quite scattered.

\begin{deluxetable}{|c|ccccccc|}[hbt]
\centerwidetable
\tabletypesize{\footnotesize}
\tablehead{\colhead{\textbf{Parameter}} & \colhead{$\beta_0$} & \colhead{$\beta_1$}& \colhead{$\beta_2$}& \colhead{$\beta_3$}& \colhead{$\beta_4$}& \colhead{$\beta_5$}& \colhead{$\beta_6$}}
\startdata
\texttt{ABSNJZH} &  -3.780e-04 & -1.992e-04 &  2.772e-04 & -1.732e-05 &
  4.523e-07 & -5.451e-09 &  2.473e-11                                        \\ \hline
\texttt{EPSX}    &  2.147e-01 & -5.051e-02 &  5.401e-03 & -1.911e-04 &
  3.148e-06 & -2.519e-08 &  7.959e-11   \\ \hline
\texttt{EPSY}    &  -9.715e-03 & 2.231e-03 & -1.492e-04 & 4.212e-06 & -6.138e-08 & 4.579e-10 & -1.367e-12   \\ \hline
\texttt{EPSZ}    &  -1.955e-02 &  2.224e-03 &  1.086e-04 & -8.265e-06 &  2.109e-07 & -2.428e-09 &  1.045e-11 \\ \hline
\texttt{MEANALP} &  2.747e-03 &  5.508e-04 &  3.417e-05 & -3.986e-06 & 1.217e-07 & -1.557e-09 &  7.261e-12                           \\ \hline
\texttt{MEANGAM} & 2.700e-03 & -4.375e-04 &  4.165e-06 &  2.204e-07 & -7.620e-09 &  9.435e-11 & -4.099e-13                           \\ \hline
\texttt{MEANGBH} &  2.332e-03 & -4.772e-04 &  2.407e-05 & -9.089e-07 &
  1.768e-08 & -1.620e-10 &  5.655e-13    \\ \hline
\texttt{MEANGBT} &  1.657e-03 & -5.667e-04 &  5.370e-05 & -2.051e-06 &
  3.661e-08 & -3.163e-10 &  1.076e-12         \\ \hline
\texttt{MEANGBZ} &  2.270e-03 & -7.084e-04 & 6.245e-05 & -2.311e-06 &
  4.112e-08 & -3.602e-10 & 1.260e-12      \\ \hline
\texttt{MEANJZD} &  1.063e-02 &  -2.389e-03 &   1.861e-04 &  -8.809e-06 & 
  2.121e-07 &  -2.411e-09 &  1.036e-11                    \\ \hline
\texttt{MEANJZH} &  1.025e-02 & -1.240e-03 &  1.574e-04 & -8.281e-06 &
  2.030e-07 & -2.343e-09 &  1.029e-11                         \\ \hline
\texttt{MEANPOT} & 4.287e-02 & -8.872e-03 &  6.295e-04 & -2.167e-05 &
  3.868e-07 & -3.458e-09 & 1.229e-11 \\ \hline
\texttt{MEANSHR} &  -3.996e-04 & 2.581e-04 & -3.342e-05 &  1.280e-06 &
 -2.422e-08 &  2.267e-10 & -8.214e-13                               \\ \hline
\texttt{R\_VALUE} & 9.112e-03 & -2.176e-03 &  1.942e-04 & -6.922e-06 &
  1.240e-07 & -1.117e-09 & 4.032e-12  \\ \hline
\texttt{SAVNCPP} & -2.819e-02 & 4.946e-03 & -1.587e-04 & -9.472e-07 &
  1.276e-07 & -2.131e-09 &  1.108e-11   \\ \hline
\texttt{SHRGT45} &  -1.564e-02 &  4.008e-03 & -3.019e-04 &  1.006e-05 &
 -1.762e-07 &  1.584e-09 & -5.730e-12    \\ \hline
\texttt{TOTBSQ}  &  -1.975e-02 & 3.703e-03 & -1.876e-04 &  4.892e-06 &
 -6.384e-08 & 3.555e-10 & -4.791e-13                \\ \hline
\texttt{TOTFX}  &   2.451e-01 & -5.886e-02 & 6.093e-03 & -2.029e-04 &
  3.164e-06 & -2.440e-08 & 7.576e-11           \\ \hline
\texttt{TOTFY}   &  -2.346e-02 & 4.590e-03 & -2.393e-04 &  5.621e-06 &
 -6.687e-08 &  3.742e-10 & -6.807e-13           \\ \hline
\texttt{TOTFZ}   &  -2.945e-02 &  3.581e-03 &  1.099e-04 & -1.064e-05 &
  3.034e-07 & -3.800e-09 & 1.740e-11            \\ \hline
\texttt{TOTPOT}  &  2.206e-02 & -5.202e-03 & 5.040e-04 & -1.980e-05 &
  3.837e-07 & -3.693e-09 &  1.415e-11       \\ \hline
\texttt{TOTUSJH} &  -1.6339e-02  & 2.592e-03 & -5.808e-05 & -5.554e-07 &
  4.066e-08 & -6.017e-10 &  2.950e-12               \\ \hline
\texttt{TOTUSJZ} &  -1.901e-02 &  3.218e-03 & -9.366e-05 &  4.323e-07 &
  2.494e-08 & -4.716e-10 &  2.530e-12                 \\ \hline
\texttt{USFLUX}  & -1.963e-02 &  3.462e-03 & -1.261e-04 &  2.062e-06 &
 -1.715e-09 & -3.156e-10 & 2.322e-12                 \\ \hline
\enddata
\caption{The constants for each 7th-degree polynomial fit for each magnetic field parameter. The fits are defined mathematically in the following way: $y = 1+\beta_0\theta+\beta_1\theta^2+\beta_2\theta^3+\beta_4\theta^5+\beta_5\theta^6+\beta_6\theta^7$, where $\beta_n$ are constants, $\theta$ is the observed heliocentric angle, and $y$ is the correction factor}
\label{tab:constants}
\end{deluxetable}

\subsection{Machine Learning Models, Feature Selection, and Hyperparameter Tuning} \label{sec:MLModels}
Machine learning (ML) has emerged as a powerful tool in heliophysics for forecasting solar transient events. In this study, we apply ML techniques to predict the occurrence of M- or X-class flares from an AR within a subsequent 24 hr time window. Known as a binary classification task, we provide positive (flare) and negative (no flare) predictions for individual point-in-time data. In ML, there are three components to every forecasting model: a training (plus validation) data set, a testing data set, and the algorithm itself, which generally includes hyperparameters, and occasionally weights or biases \citep[e.g., see][for more details]{badillo2020introduction}. In this work, we use every possible combination of the temporally segmented partitions of SWAN-SF to train/test our algorithm (1/2, 1/3, 1/4, 1/5, 2/1, etc.). This was done for both corrected and uncorrected data. To address the problem of differing scales within the derived magnetogram parameters, we renormalize each training dataset feature distribution to a mean of 0 and a variance of 1. This transformation is calculated by using the following formula: $z = \frac{x-u}{s}$, where $z$ is the transformed value, $x$ is the input value, $u$ is the mean of the training samples, and $s$ is the standard deviation of the training samples. We then apply this exact transformation to the testing dataset, with $u$ and $s$ being the (same) values obtained during the scaling of the training set. By implementing this training/testing approach, we obtain a wealth of results critical to judging the effectiveness of corrections. 

We selected the support vector machine (SVM) algorithm for forecasting purposes. SVM is a robust, efficient, and widely used classifier, making it an excellent choice for flare prediction. It has also been shown to perform comparably to decision tree and multilayer perceptron forecasters when utilizing point-in-time magnetogram data, providing superior performance, while not compromising too much in terms of complexity and training time \citep{Goodwin_Sadykov_Martens_2024}. This algorithm aims to develop a multidimensional hyperplane that maximizes the separation between labels within a feature space. In the context of this paper, our labels are the flaring/non-flaring events, and our feature space consists of some subset of the 264 descriptive magnetogram parameters. More often than not, labels are not linearly separable, requiring a more sophisticated hyperplane to be developed. As a remedy, ``kernel tricks" are frequently applied to remap the feature space into a higher dimension. In this work, we utilize the popular radial basis function (RBF) kernel, which is dependent on two hyperparameters: $C$ and $\gamma$. Both influence the complexity of the final decision boundary, with $C$ behaving as the penalty parameter for misclassification and $\gamma$ as the ``width" of the kernel. The larger $C$ and $\gamma$ are, the more complicated the resulting decision boundary will be \citep[see][for more details]{bobra2015solar, Goodwin_Sadykov_Martens_2024}. The RBF kernel is defined by the following formula: $K(x,x^\prime)=exp(-\gamma||x-x^\prime||^2)$, where $x$ and $x^\prime$ are feature vectors \citep{ding2021random}. The kernel, in essence, calculates the Euclidian distance between two points, mapping the data into a higher dimensional parameter space to better capture the similarity between points.


To ensure that we achieve clear-cut results on the impacts of projection effects on forecasting, we structure our feature selection methodology to amplify potential improvements from corrections. We calculate the Fisher score (F-score) of the 264 descriptive magnetogram parameters for all corrected and uncorrected training partitions. The F-score is mathematically defined in the following way \citep{chen2006combining,guFscore}: 
\begin{equation}
    F(i) = \frac{(\bar{x}_i^+-\bar{x}_i)^2+(\bar{x}_i^--\bar{x}_i)^2}{\frac{1}{n^+-1}\sum_{c=1}^{n^+}(\bar{x}_{c,i}^+-\bar{x}_i)^2+\frac{1}{n^--1}\sum_{c=1}^{n^-}(\bar{x}_{c,i}^--\bar{x}_i)^2}
\end{equation}
where, for a given feature $i$, $\bar{x}_i^+$ is the average value across flaring events, $\bar{x}_i^-$ is the average value across non-flaring events, $\bar{x}_i$ is the average value across all events, $n^+$ is the number of flaring events, and  $n^-$ is the number of non-flaring events. Features with large F-scores have flaring and non-flaring distributions that are highly separated with small standard deviations, meaning they are useful to include when making predictions. Once we calculate the F-scores within a given partition, we rank the parameters from highest to lowest depending on their score. However, we only include parameters that show an improvement to their F-score after corrections were applied. For example, when we run this methodology on partition 1, we find that \texttt{SAVNCPP\_max} (the maximum value of the sum of the absolute value of the net current per polarity) achieves an F-score of 16,920.76 and 17,220.60 for the uncorrected and corrected versions respectively. This means we will include this parameter in the rankings. However, \texttt{ABSNJZH\_max} (the max value of the absolute value of the net current helicity) achieves an F-score of 18,327.77 and 17,591.47 for the uncorrected and corrected versions respectively. Thus, we do not include this parameter in the rankings. While this may not reflect the most optimal feature selection methodology, this logic ensures that potential improvements from corrections do not go unnoticed. In Section \ref{sec:doCorrectionsImproveForecasting}, we explore an additional selection methodology that includes parameters with no improvement.

After getting our list of improved parameters, we select the 25 with the highest F-score to use in both our uncorrected and corrected data sets. Previously it was found that for point-in-time magnetogram forecasting, the number of features (when selected based on F-score) has minimal impact on the resulting skill scores \citep{Goodwin_Sadykov_Martens_2024}. Thus, for this study, we settled on 25 features as this balances both performance and training time. This process was repeated for every new training partition used. Please see Table \ref{tab:features} for the complete list of features selected.

\begin{longrotatetable}
\begin{deluxetable}{|cl|cl|cl|cl|cl|}
\centerwidetable
\tabletypesize{\tiny}
\tablecaption{The 25 parameters selected for each training partition. Parameters were only chosen if they improved their F-scores after corrections were applied. The rank highlights the parameter's original order when all parameters were ranked by the highest F-score they achieved across both the corrected and uncorrected data. The relative improvement to the F-score after corrections were applied is shown in parentheses.}
\tablehead{\multicolumn{2}{c}{\textbf{Partition 1}} &
  \multicolumn{2}{c}{\textbf{Partition 2}} &
  \multicolumn{2}{c}{\textbf{Partition 3}} &
  \multicolumn{2}{c}{\textbf{Partition 4}} &
  \multicolumn{2}{c}{\textbf{Partition 5}} \\  \colhead{\textbf{Rank}} & \colhead{\textbf{Parameter}} & \colhead{\textbf{Rank}} & \colhead{\textbf{Parameter}} & \colhead{\textbf{Rank}} & \colhead{\textbf{Parameter}} & \colhead{\textbf{Rank}} & \colhead{\textbf{Parameter}} & \colhead{\textbf{Rank}} & \colhead{\textbf{Parameter}}}
\startdata
3 & 
\texttt{SAVNCPP\_max} (1.8\%) & 
16 & \texttt{ABSNJZH\_gderivative\_stddev} (0.3\%) & 
31 & \texttt{TOTUSJH\_stddev} (7.6\%) & 
1 & \texttt{TOTUSJH\_median} (0.002\%) & 
17 & \texttt{TOTUSJZ\_last\_value} (1.8\%) \\ \hline
4 &
  \texttt{SAVNCPP\_last\_value} (1.1\%) &
  32 &
  \texttt{USFLUX\_last\_value} (3.6\%) &
  33 &
  \texttt{USFLUX\_last\_value} (0.4\%) &
  2 &
  \texttt{TOTUSJH\_min} (0.1\%) &
  27 &
  \texttt{TOTUSJZ\_max} (0.5\%) \\ \hline
7 &
  \texttt{SAVNCPP\_mean} (1.0\%) &
  33 &
  \texttt{USFLUX\_median} (2.5\%) &
  35 &
  \texttt{USFLUX\_mean} (0.4\%) &
  3 &
  \texttt{TOTUSJH\_mean} (0.02\%) &
  33 &
  \texttt{SAVNCPP\_min} (0.5\%) \\ \hline
8 &
  \texttt{SAVNCPP\_median} (0.8\%) &
  34 &
  \texttt{USFLUX\_mean} (2.6\%) &
  36 &
  \texttt{USFLUX\_median} (0.2\%) &
  10 &
  \texttt{TOTUSJZ\_min} (0.6\%) &
  38 &
  \texttt{TOTUSJH\_stddev} (5.3\%) \\ \hline
9 &
  \texttt{ABSNJZH\_stddev} (0.2\%) &
  35 &
  \texttt{USFLUX\_max} (3.8\%) &
  37 &
  \texttt{USFLUX\_max} (0.8\%) &
  11 &
  \texttt{TOTUSJZ\_median} (0.4\%) &
  43 &
  \texttt{TOTFY\_stddev} (1.2\%) \\ \hline
10 &
  \texttt{SAVNCPP\_min} (0.7\%) &
  36 &
  \texttt{USFLUX\_min} (1.4\%) &
  44 &
  \texttt{TOTFY\_stddev} (3.9\%) &
  13 &
  \texttt{TOTUSJZ\_mean} (0.4\%) &
  54 &
  \texttt{TOTUSJZ\_stddev} (11.5\%) \\ \hline
13 &
  \texttt{SAVNCPP\_stddev} (2.8\%) &
  37 &
  \texttt{TOTUSJH\_stddev} (15.1\%) &
  47 &
  \texttt{MEANPOT\_min} (10.5\%) &
  17 &
  \texttt{USFLUX\_last\_value} (1.3\%) &
  57 &
  \texttt{MEANPOT\_min} (12.6\%) \\ \hline
14 &
  \texttt{TOTUSJH\_median} (1.1\%) &
  38 &
  \texttt{TOTPOT\_median} (0.02\%) &
  49 &
  \texttt{R\_VALUE\_min} (5.1\%) &
  21 &
  \texttt{USFLUX\_mean} (1.1\%) &
  58 &
  \texttt{TOTPOT\_mean} (6.9\%) \\ \hline
15 &
  \texttt{TOTUSJH\_max} (1.1\%) &
  40 &
  \texttt{TOTUSJH\_var} (24.9\%) &
  50 &
  \texttt{USFLUX\_stddev} (16.9\%) &
  22 &
  \texttt{USFLUX\_median} (0.9\%) &
  59 &
  \texttt{R\_VALUE\_min} (1.8\%) \\ \hline
16 &
  \texttt{TOTUSJH\_mean} (1.0\%) &
  41 &
  \texttt{TOTBSQ\_stddev} (13.9\%) &
  51 &
  \texttt{TOTUSJZ\_stddev} (18.8\%) &
  24 &
  \texttt{USFLUX\_max} (3.6\%) &
  60 &
  \texttt{TOTUSJZ\_gderivative\_stddev} (1.1\%) \\ \hline
19 &
  \texttt{TOTUSJH\_min} (0.5\%) &
  43 &
  \texttt{TOTFZ\_min} (8.6\%) &
  53 &
  \texttt{TOTUSJH\_var} (3.3\%) &
  31 &
  \texttt{TOTUSJH\_max} (1.7\%) &
  61 &
  \texttt{USFLUX\_stddev} (22.0\%) \\ \hline
25 &
  \texttt{ABSNJZH\_gderivative\_stddev} (0.9\%) &
  44 &
  \texttt{TOTFZ\_last\_value} (9.7\%) &
  57 &
  \texttt{TOTFY\_var} (16.5\%) &
  36 &
  \texttt{TOTBSQ\_stddev} (23.4\%) &
  62 &
  \texttt{SHRGT45\_min} (28.6\%) \\ \hline
27 &
  \texttt{SAVNCPP\_gderivative\_stddev} (0.4\%) &
  45 &
  \texttt{TOTPOT\_mean} (3.7\%) &
  58 &
  \texttt{SHRGT45\_min} (22.1\%) &
  38 &
  \texttt{TOTFZ\_min} (4.9\%) &
  63 &
  \texttt{SHRGT45\_last\_value} (32.0\%) \\ \hline
30 &
  \texttt{TOTUSJZ\_last\_value} (1.9\%) &
  47 &
  \texttt{TOTFZ\_mean} (9.1\%) &
  59 &
  \texttt{R\_VALUE\_mean} (3.0\%) &
  40 &
  \texttt{ABSNJZH\_gderivative\_stddev} (1.4\%) &
  64 &
  \texttt{SHRGT45\_median} (29.7\%) \\ \hline
31 &
  \texttt{TOTPOT\_last\_value} (3.5\%) &
  48 &
  \texttt{TOTFZ\_median} (9.2\%) &
  62 &
  \texttt{R\_VALUE\_last\_value} (2.0\%) &
  41 &
  \texttt{TOTFZ\_last\_value} (2.9\%) &
  65 &
  \texttt{SHRGT45\_mean} (29.6\%) \\ \hline
32 &
  \texttt{TOTUSJZ\_min} (3.5\%) &
  50 &
  \texttt{USFLUX\_var} (89.5\%) &
  63 &
  \texttt{R\_VALUE\_median} (2.3\%) &
  42 &
  \texttt{TOTFZ\_mean} (2.3\%) &
  66 &
  \texttt{MEANGAM\_min} (45.9\%) \\ \hline
33 &
  \texttt{TOTUSJZ\_median} (3.4\%) &
  51 &
  \texttt{USFLUX\_stddev} (29\%) &
  64 &
  \texttt{SHRGT45\_mean} (22.7\%) &
  43 &
  \texttt{TOTFZ\_median} (1.9\%) &
  67 &
  \texttt{MEANSHR\_min} (18.1\%) \\ \hline
34 &
  \texttt{TOTUSJZ\_mean} (3.6\%) &
  52 &
  \texttt{TOTFZ\_max} (8.6\%) &
  65 &
  \texttt{SHRGT45\_median} (22.6\%) &
  49 &
  \texttt{USFLUX\_stddev} (36.9\%) &
  68 &
  \texttt{MEANSHR\_last\_value} (22.7\%) \\ \hline
35 &
  \texttt{TOTUSJH\_stddev} (9.2\%) &
  55 &
  \texttt{TOTBSQ\_gderivative\_stddev} (1.2\%) &
  66 &
  \texttt{SHRGT45\_last\_value} (23.8\%) &
  50 &
  \texttt{TOTFY\_last\_value} (1.2\%) &
  69 &
  \texttt{MEANSHR\_mean} (20.1\%) \\ \hline
36 &
  \texttt{TOTUSJZ\_max} (2.6\%) &
  57 &
  \texttt{TOTBSQ\_var} (55.7\%) &
  67 &
  \texttt{MEANSHR\_min} (14.5\%) &
  51 &
  \texttt{TOTFY\_median} (0.9\%) &
  70 &
  \texttt{MEANSHR\_median} (20.2\%) \\ \hline
43 &
  \texttt{TOTBSQ\_stddev} (4.8\%) &
  60 &
  \texttt{TOTPOT\_last\_value} (6.8\%) &
  68 &
  \texttt{MEANGAM\_min} (34.5\%) &
  52 &
  \texttt{TOTFY\_mean} (0.7\%) &
  72 &
  \texttt{USFLUX\_var} (114.5\%) \\ \hline
45 &
  \texttt{TOTUSJH\_var} (9.7\%) &
  61 &
  \texttt{TOTUSJZ\_stddev} (10.9\%) &
  69 &
  \texttt{MEANSHR\_mean} (15.8\%) &
  53 &
  \texttt{TOTFY\_max} (0.4\%) &
  74 &
  \texttt{MEANGAM\_last\_value} (59.3\%) \\ \hline
46 &
  \texttt{TOTFY\_gderivative\_stddev} (3.0\%) &
  63 &
  \texttt{R\_VALUE\_min} (8.2\%) &
  70 &
  \texttt{MEANSHR\_median} (15.9\%) &
  54 &
  \texttt{MEANPOT\_min} (2.0\%) &
  75 &
  \texttt{MEANGAM\_median} (56.7\%) \\ \hline
51 &
  \texttt{TOTBSQ\_gderivative\_stddev} (1.3\%) &
  65 &
  \texttt{R\_VALUE\_mean} (6.3\%) &
  71 &
  \texttt{MEANSHR\_last\_value} (17.1\%) &
  56 &
  \texttt{TOTFZ\_gderivative\_stddev} (83.5\%) &
  76 &
  \texttt{MEANGAM\_mean} (56.7\%) \\ \hline
53 &
  \texttt{TOTPOT\_max} (11.1\%) &
  66 &
  \texttt{R\_VALUE\_last\_value} (5.6\%) &
  73 &
  \texttt{MEANGAM\_median} (41.0\%) &
  57 &
  \texttt{R\_VALUE\_min} (6.7\%) &
  79 &
  \texttt{SHRGT45\_max} (33.8\%) \\ \hline
\enddata
\label{tab:features}
\end{deluxetable}
\end{longrotatetable}

Finally, to ensure that each ML model provides reasonable predictive performance, an exhaustive hyperparameter grid search was applied to all training partitions (both corrected and uncorrected) to optimize each model's $C$ and $\gamma$ hyperparameter. For a given training partition, a model was generated for every possible combination of the following values for $C$ (0.0001, 0.001, 0.01, 0.1, 1, 10, 100, 1000) and $\gamma$ (``scale", 0.0001, 0.001, 0.01, 0.1, 1, 10, 100, 1000). Where $\gamma$ = ``scale" utilizes the inverse of the number of features times the variance of the feature vector. For each of these models, the \texttt{class\_weight} hyperparameter was set to \texttt{"balanced"} (meaning no under/oversampling techniques were applied), and the \texttt{kernel} parameter was set to \texttt{"rbf"}\footnote{See \href{https://scikit-learn.org/stable/modules/generated/sklearn.svm.SVC.html}{https://scikit-learn.org/stable/modules/generated/sklearn.svm.SVC.html} for more details.}. These models were then trained/tested using a stratified group three-fold cross-validation. This ensures that within the training and testing folds, no data overlaps between ARs, and a similar number of flaring and non-flaring events are present. The model that produced the highest true skill statistic score (see Section \ref{sec:performanceMetrics}) was then selected for application to the full training dataset. The final hyperparameters for each training partition are listed in Table \ref{tab:finalizedHyperparameters}.

\begin{deluxetable}{|c|c|c|}
\tabletypesize{\footnotesize}
\tablehead{\colhead{\textbf{Partition}} & \colhead{\textbf{Uncorrected}} & \colhead{\textbf{Corrected}}}
\startdata
1 & $C$: 10, $\gamma$: 0.0001 & $C$: 0.01, $\gamma$: 0.01\\ \hline
2 & $C$: 0.01, $\gamma$: ``scale" & $C$: 0.1, $\gamma$: 0.0001\\ \hline
3 & $C$: 100, $\gamma$: 0.0001 & $C$: 10, $\gamma$: 0.0001\\ \hline
4 & $C$: 0.001, $\gamma$: 0.01 & $C$: 0.0001, $\gamma$: ``scale"\\ \hline
5 & $C$: 0.001, $\gamma$: 0.01 & $C$: 0.1, $\gamma$: 0.0001\\ \hline
\enddata
\caption{The final combination of $C$ and $\gamma$ for each training partition for the uncorrected and corrected data sets. The ``scale" parameter for $\gamma$ utilizes the inverse of the number of features times the variance of the feature vector.}
\label{tab:finalizedHyperparameters}
\end{deluxetable}

\subsection{Performance Metrics} \label{sec:performanceMetrics}
To analyze the performance of our models, we focus on four different metrics: true positive rate (TPR), false positive rate (FPR), true skill statistic (TSS), and Heidke skill score (HSS$_2$). They are each defined mathematically in the following way:
\begin{equation}
    TPR = \frac{TP}{TP+FN},
\end{equation}
\begin{equation}
    FPR = \frac{FP}{FP+TN},
\end{equation}
\begin{equation}
    TSS = \frac{TP}{TP+FN} - \frac{FP}{FP+TN},
\end{equation}
\begin{equation}
    HSS_2 = \frac{2\times[(TP\times TN)-(FN\times FP)]}{(TP+FN)\times(FN+TN)+(TP+FP)\times(FP+TN)},
\end{equation}
where TP are true positives (the number of correctly predicted flaring events), TN are true negatives (the number of correctly predicted non-flaring events), FP are false positives (the number of non-flaring events predicted as flaring), and FN are false negatives (the number of flaring events predicted as non-flaring). The TPR can be thought of as the likelihood of correctly predicting that a flaring event will occur, while the FPR is the likelihood of predicting that a flaring event will occur when it does not. Taking the difference of these metrics gives us the TSS score. This metric ranges from -1 to +1, and provides insight into the improvement of a classifier over a random forecast. A score of 0 would be given to a classifier that provides purely positive, negative, or simply random forecasts. A score of +1 would be given to a classifier that is always right, and a score of -1 would be given to a classifier that is always wrong. This metric is frequently used in flare forecasting as it is impervious to class imbalance \citep{bobra2015solar}. Lastly, similar to TSS, HSS$_2$ also provides insight into a model's performance improvement over a random forecast with scores ranging from -1 to +1. However, unlike TSS, HSS$_2$ is highly susceptible to class imbalance, with scores decreasing when the imbalance increases (see Figure 2 in \citealt{bobra2015solar} and Figure 4 in \citealt{Ahmadzadeh_Aydin_Georgoulis_Kempton_Mahajan_Angryk_2021}). Due to the nature of our data set, changes in the number of false positives and true negatives have a much greater impact on this score. Thus, it's important to include both TSS and HSS$_2$ when analyzing performance to see the results holistically. If the reader would like to learn more about these metrics in the context of flare forecasting, please see \cite{bloomfield2012toward}, \cite{bobra2015solar}, \cite{Ahmadzadeh_Aydin_Georgoulis_Kempton_Mahajan_Angryk_2021}, and \cite{ Goodwin_Sadykov_Martens_2024}.
\section{Results and Discussion} \label{sec:results}
In the following sections, we discuss how corrections change forecasts at different heliocentric angles (Section \ref{sec:whatAngleAreImprovements}), how performance metrics change after corrections are applied (see Section \ref{sec:doCorrectionsImproveForecasting}), and what are some other potential methods that can be used to address projection effects (see Section \ref{sec:otherPotential}).

\subsection{At What Heliocentric Angles Do Corrections ``Work"?}
\label{sec:whatAngleAreImprovements}
\begin{figure}[htb]
    \centering
    \includegraphics[width=\textwidth]{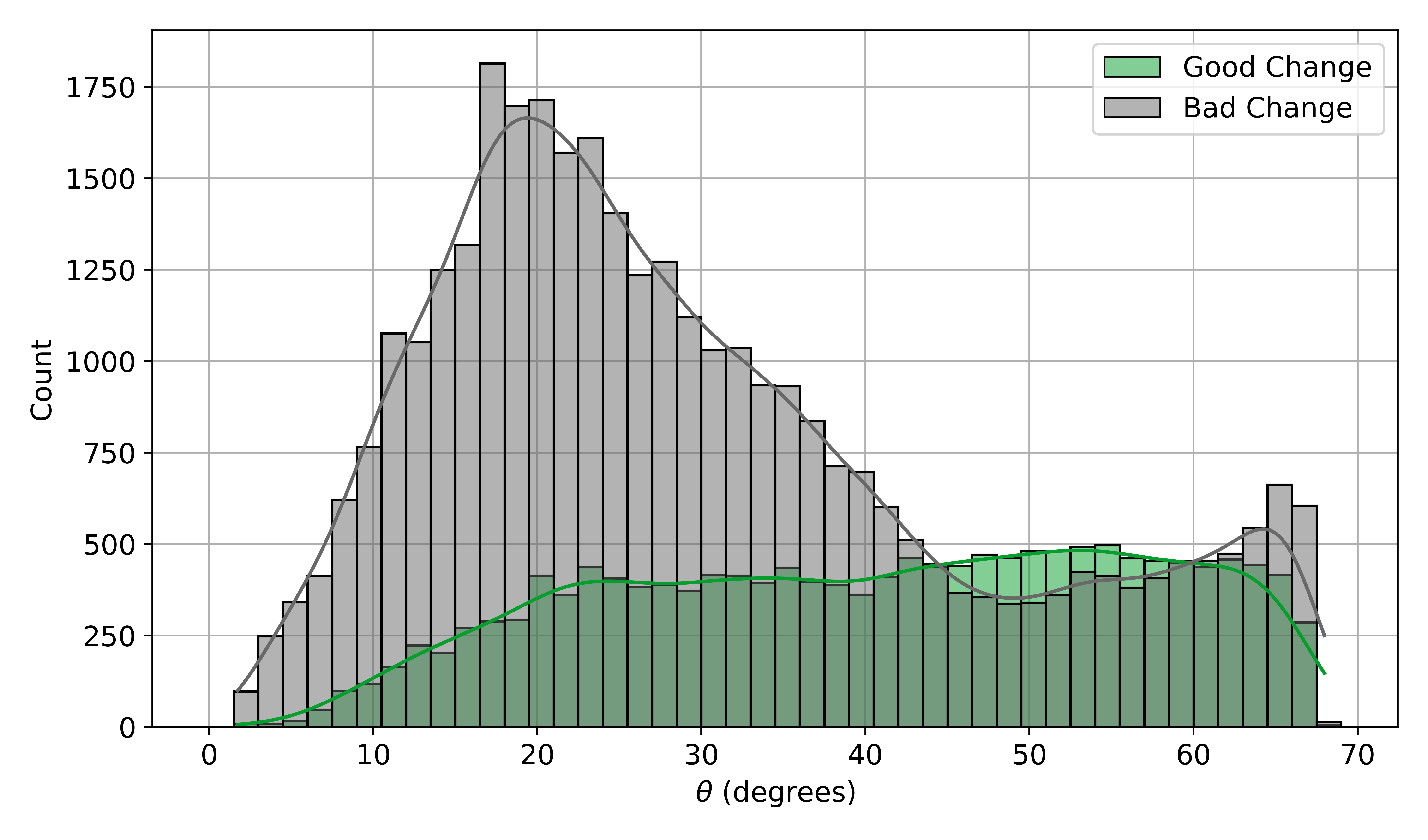}
    \caption{A histogram illustrating how removing projection effects impacts the predictions made at different average heliocentric angles ($\theta$) across all training/testing partition combinations. Bars are color-coded to indicate if an incorrect forecast was corrected (Good Change) or if a correct forecast became incorrect (Bad Change) after projection effects were removed. A solid line representing the kernel density estimation of each distribution is plotted as well.}
    \label{fig:angleHistogram}
\end{figure}

Gathering the results across all training/testing partition pairs allows us to evaluate the ``good changes" and ``bad changes" made to our forecasts following corrections. In this context, a ``good change" equates to an originally incorrect forecast that was later corrected after removing projection effects. In contrast, a ``bad change" is an originally correct forecast that is now incorrect after removing projection effects. In Figure \ref{fig:angleHistogram}, we highlight these changes as a function of the mean $\theta$. Events whose predictions stayed the same between forecasts were not included. It is clear from the figure that the ``bad changes" significantly outnumber the ``good changes" near the central meridian. However, as we move toward the limb, they quickly balance out. Using the intersection of the kernel density estimation \citep[see][for more details]{weg,Waskom2021} for each distribution, we find that this occurs at $\approx$44.5$^\circ$. This suggests two things: first, predicting with corrective data is more effective near the limb. Second, if we want to maximize performance, an ensemble-like forecaster must be applied. We suggest using the uncorrected data/classifier for predictions of ARs with $\theta$s less than 44.5$^\circ$ and the corrected data/classifier for anything above this angle. This is explored further in Section \ref{sec:doCorrectionsImproveForecasting}.

\begin{figure}[htb]
    \centering
    \includegraphics[width=\textwidth]{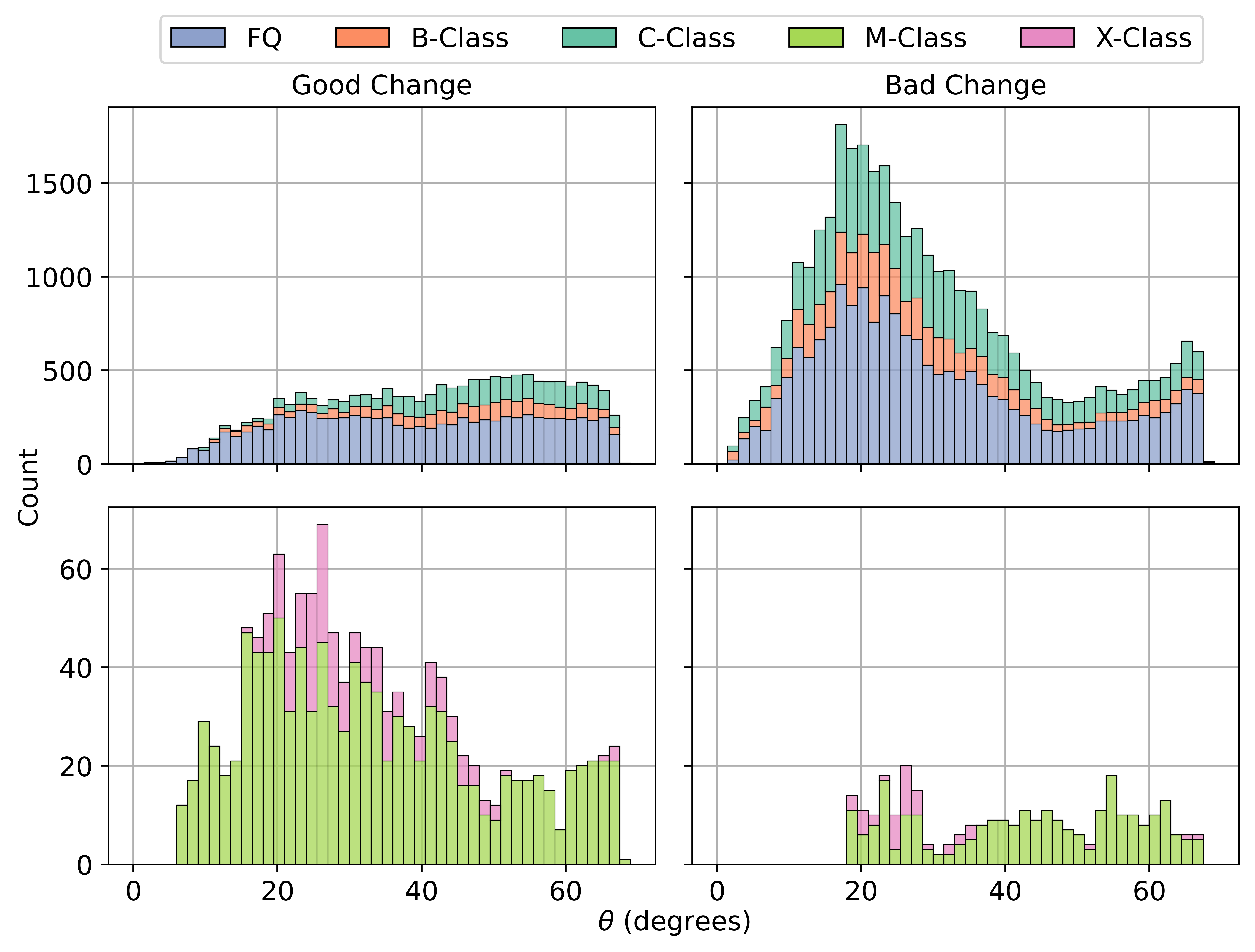}
    \caption{A stacked histogram highlighting the ``good" (left column) and ``bad" (right column) changes to forecast after removing projection effects as a function of heliocentric angle ($\theta$). These data come from the collection training/testing partition experiments. Results are color-coded based on flaring class. Flare quiet (FQ), B-, and C-class events are shown in the top panels. M- and X-class events are shown in the bottom panels.}
    \label{fig:angleHistogramClass}
\end{figure}

In addition to looking at the data as a whole, we can dive into the flaring classes themselves to see those most affected by corrections. In Figure \ref{fig:angleHistogramClass}, we plot the same results from Figure \ref{fig:angleHistogram}, however, the data is now separated into the ``good" (left column) and ``bad" (right column) changes and color-coded based on the non-flaring (top row) and flaring (bottom row) events that occur at a specific mean $\theta$. From this plot, we can see that the M- and X-class flares tend to have more ``good" than ``bad" changes after corrections are applied, with the opposite observed for the non-flaring data. Thus, we can expect to see increases in both the true positive and false positive rates when analyzing the performance metrics in Section \ref{sec:doCorrectionsImproveForecasting}. However, we expect this increase to be quite minimal. This is because the changes in our predictions occur for only a small subset of our data. Analyzing the results across every possible partition combination we find that, on average, ``good changes" occur to the flaring data around 5.1\% $\pm$ 6.8\% of the time, with ``bad changes" around 1.3\% $\pm$ 2.1\%. For the non-flaring data, ``good changes" occur around 1.1\% $\pm$ 1.0\% of the time, with ``bad changes" around 2.7\% $\pm$ 1.4\%. Unfortunately, there is no clear explanation for why these changes occur the way they do. However, we speculate that an improvement in M- and X-class forecasting may be due to the activity of the selected ``benchmark" ARs. We find that around 37\% of these ARs produce at least one M- or X-class flare during their lifetime. Additionally, these ARs make up 51\% of the total number of flaring ARs within SWAN-SF. Since we have such a large representation of flaring ARs when calibrating our corrections, this could be a reasonable explanation for the phenomenon observed here.
\subsection{Analysis Of Performance Metrics For Each Training-Testing Partition}
\label{sec:doCorrectionsImproveForecasting}
\begin{figure}[htb]
    \centering
    \includegraphics[width=\textwidth]{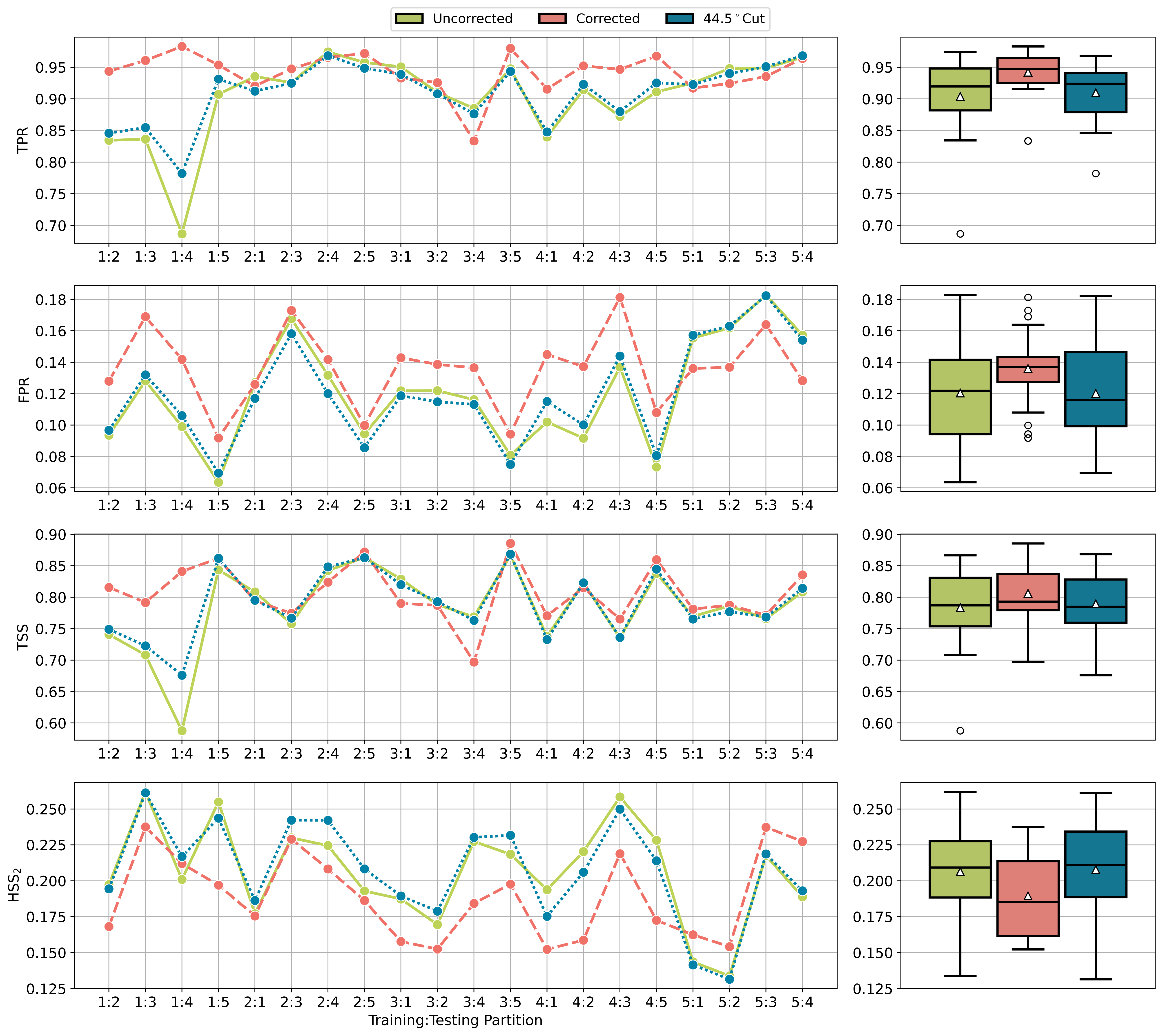}
    \caption{The true positive rate (TPR), false positive rate (FPR), true skill statistic (TSS), and Heidke skill scores (HSS$_2$) for every training and testing partition pair from SWAN-SF (denoted training:testing). Results are color-coded to indicate whether the classifier was trained and tested with uncorrected data (green), corrected (coral), or a combination of the two (blue - “44.5° Cut”). The “44.5° Cut” results use the uncorrected forecasts below heliocentric angles of 44.5° and the corrected forecasts above this angle. Within the box plots, the triangle represents the data's mean, while the circles highlight outliers. The outliers are defined as any value lying 1.5 times the interquartile range above the third quartile or below the first quartile.}
    \label{fig:scoresScatter}
\end{figure}
Now, let us analyze the changes to performance metrics, specifically TPR, FPR, TSS, and HSS$_2$. Figure \ref{fig:scoresScatter} highlights our results. The scatter plots on the left show the scores for each partition combination, while the box plots on the right summarize the findings across all combinations. Results are color-coded to indicate whether the classifier was trained/tested with uncorrected data (grey), corrected data (maroon), or a combination of the two (salmon pink - ``44.5$^\circ$ Cut"). As was previously mentioned, corrections appear to be the most useful beyond mean $\theta$s of 44.5$^\circ$. Thus, the ``44.5$^\circ$ Cut" results use the uncorrected forecasts below mean $\theta$s of 44.5$^\circ$ and the corrected forecasts above this angle. From the figure, we find that on average, removing projection effects increases both the TPR and FPR by a few percent. In turn, this produces an enhancement in the TSS scores but a reduction in the HSS$_2$ scores. Additionally, it is interesting to note that removing projection effects reduces the overall data spread, with the interquartile range decreasing for both the TPR and FPR. This suggests we generally get more consistent results after corrections, which is beneficial from a forecaster's perspective. Looking at the ``44.5$^\circ$ Cut", which should maximize our performance improvements, we see an ever-so-slight increase to both the average TSS and HSS$_2$ scores. 

\begin{figure}[htb]
    \centering
    \includegraphics[width=.88\textwidth]{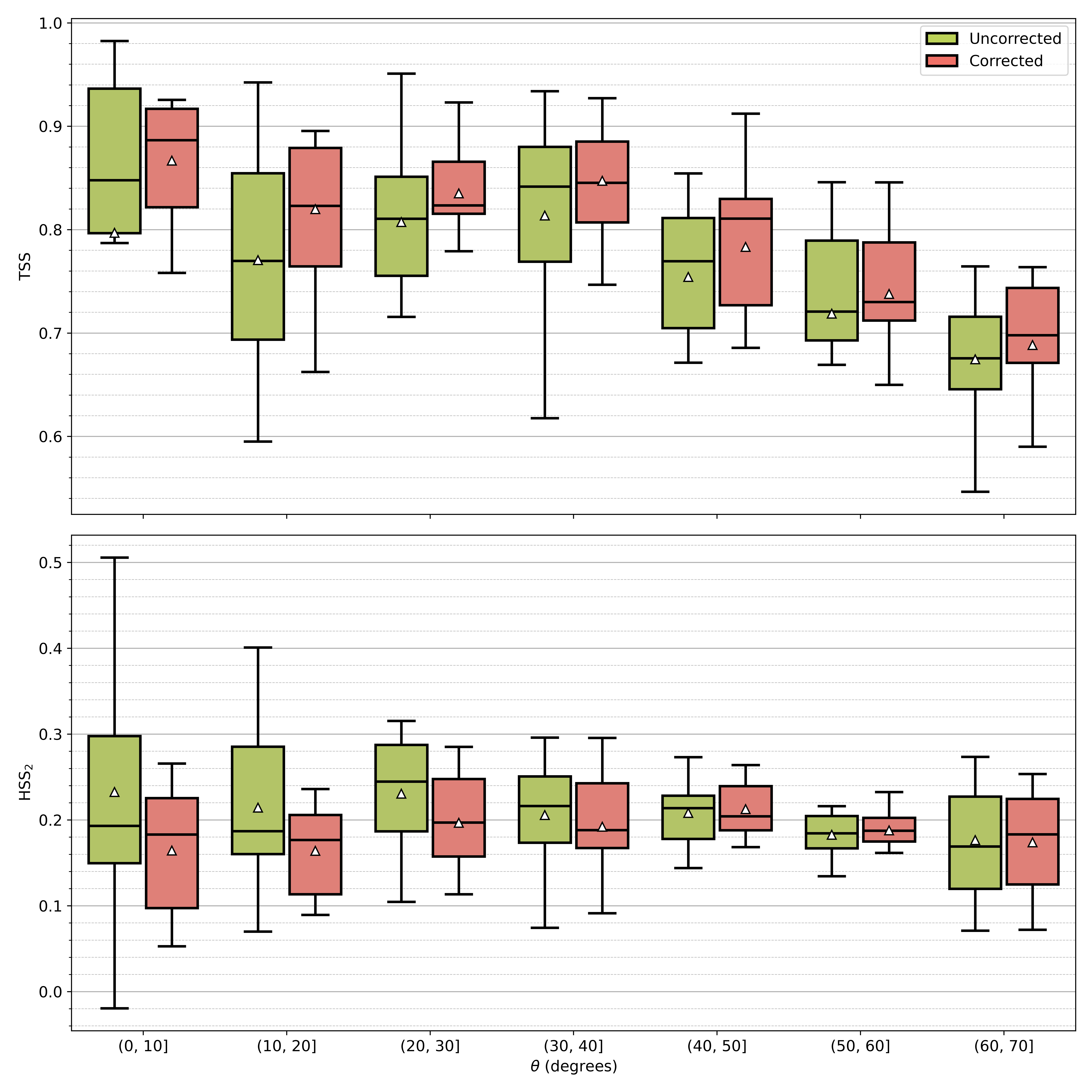}
    \caption{The TSS and HSS$_2$ scores for each training/testing partition pair for the uncorrected (green) and corrected (coral) data as a function of heliocentric angle ($\theta$). Results are grouped into 10-degree bins. The triangles indicate the mean of the data. Outliers have been removed and are not shown in the plot.}
    \label{fig:binnedHA}
\end{figure}

We want to emphasize that although we observe changes in our performance metrics when correcting our data, the hyperparameters we select to train our model can significantly impact these variations. With only a slight tweak to $C$ and $\gamma$, our findings may be completely altered, decreasing the true positive and false positive rates. This leads to a classifier with somewhat better HSS$_2$ scores but worse TSS scores after corrections. \textit{Nonetheless, the key takeaway should be that correcting for projection effects has only a small impact on the achieved skill scores.} This fact has some potentially profound consequences. 

First, it may suggest that the correction methodology utilized in this work is not sophisticated enough. If we take a look at how the TSS and HSS$_2$ scores change as a function of $\theta$ (Figure \ref{fig:binnedHA}), we see that near the central meridian, at (0, 10], TSS scores hover at $\approx$ 0.85, but slowly drop to $\approx$ 0.67 when moving toward the limb. Once corrections are applied, this drop-off still exists, albeit slightly reduced, suggesting that there is still room for improvements from our corrections. For HSS$_2$ scores, very little change is observed when approaching the limb, suggesting there is not much room for improvement here. A potential remedy to this problem could be to utilize more fundamental corrections targeting the magnetograms themselves. For example, \cite{leka2022identifying} implemented several advanced spectropolarimetric inversion techniques to remove biases within the data. It would be interesting to see if applying similar methods to the magnetograms collected for SWAN-SF would provide greater improvements to forecasting.

Second, and perhaps the least desirable interpretation of this outcome, is that the magnetogram data itself is limited in its predictive capabilities for flaring events. Given that applying scaling factors to the data set had minimal effect on the results indicates that classifiers tend to rely on very generalized associations when making predictions. This suggests that additional data sources, such as EUV images \citep{nishizuka2018deep,Alipour_Mohammadi_Safari_2019,Krista_Chih_2021,leka2023properties} and spectra \citep{Panos_Kleint_2020,Woods_Dalda_Pontieu_2021,Huwyler_Melchior_2022,Panos_Kleint_Zbinden_2023} used in recent forecasting works, need to be implemented in conjunction with magnetograms to progress the field further. 

\begin{figure}[htb]
    \centering
    \includegraphics[width=\textwidth]{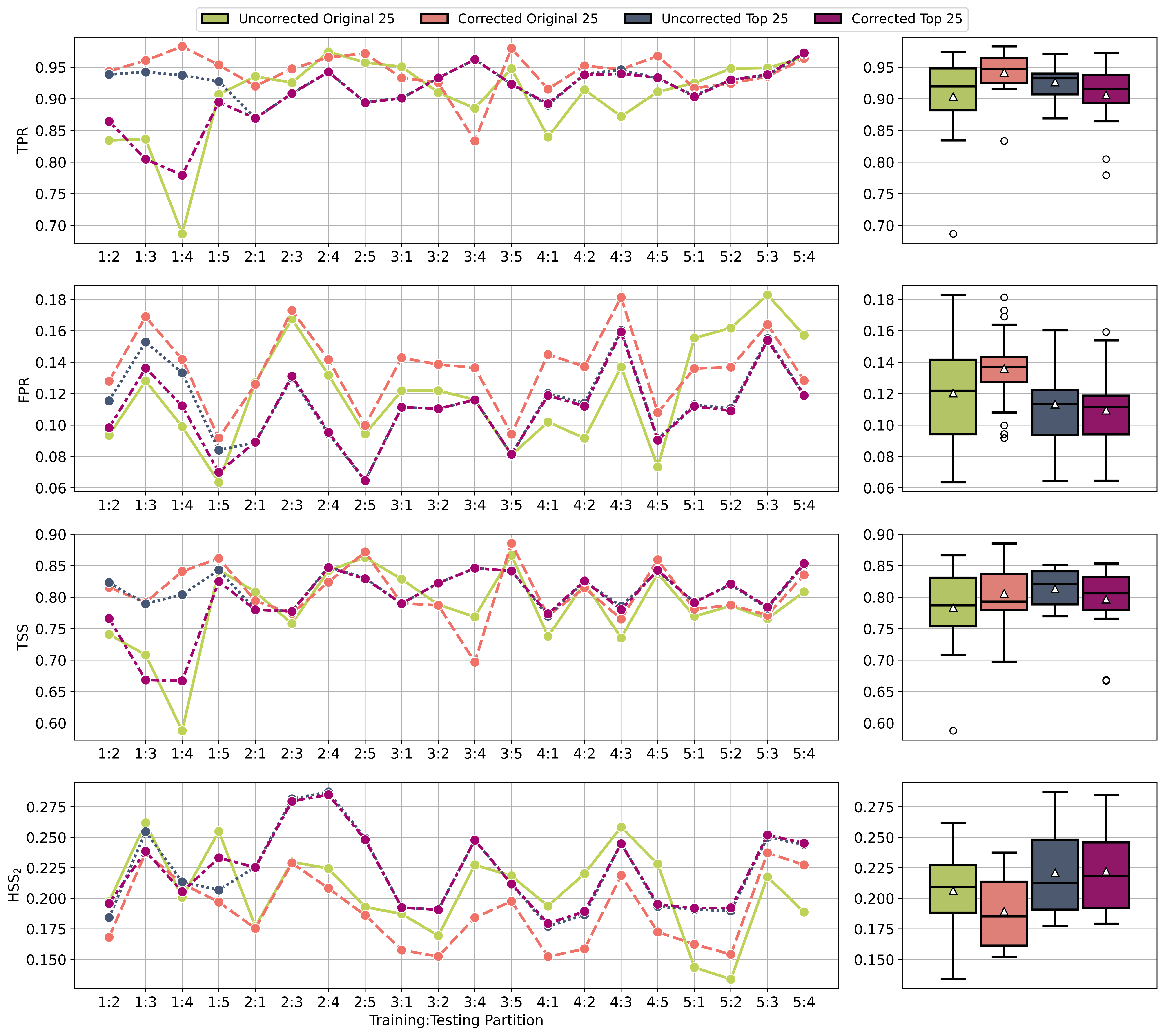}
    \caption{The true positive rate (TPR), false positive rate (FPR), true skill statistic (TSS), and Heidke skill scores (HSS$_2$) for every training/testing partition combination from SWAN-SF. Results are color-coded to indicate whether the classifier was trained/tested with uncorrected data with the original 25 features (green), the corrected data with the original 25 features (coral), the uncorrected data with the top 25 features (blue), or the corrected data with the top 25 features (purple). See the caption in Figure \ref{fig:scoresScatter} for the meaning of the triangles and circles in the box plots.}
    \label{fig:scoresCombo}
\end{figure}

As previously mentioned, our feature selection methodology is not necessarily optimized for flare forecasting. Instead, we chose parameters with the greatest potential to maximize performance improvements when applying corrections. Nonetheless, we also wanted to investigate how performance changes when selecting the most effective features for forecasting. Here, we carry out the same feature ordering procedure mentioned in Section \ref{sec:MLModels}, however, we now select all of the top 25 ranking features, not just the ones that show improvements to their F-score. Like before, the uncorrected dataset uses only the uncorrected features, however, the corrected classifier may now use some combination of corrected and uncorrected parameters. This is most relevant for partitions 1 and 4, where 12 and 10 of the chosen features, respectively, have been corrected. For partitions 2 and 5, only 1 corrected feature is used, and for partition 3, none of the corrected features are used. Our results are shown in Figure \ref{fig:scoresCombo}. Here, we see a decent discrepancy between the new uncorrected and corrected TSS scores when using partition 1 for training, however, this is not observed in partition 4 or any other partitions. This again emphasizes our point that projection effect corrections have little impact on performance, or at the very least, results seem to be dependent on the partition utilized. Comparing the use of the top 25 features versus the original feature selection methodology, we see that the performance differences are fairly minimal, with the new methodology only slightly outperforming the old. This once again reinforces the idea that there may be some serious limitations when utilizing point-in-time magnetogram data for forecasting.

\subsection{How Else Can We Consider Projection Effects?}
\begin{figure}[htb]
    \centering
    \includegraphics[width=\textwidth]{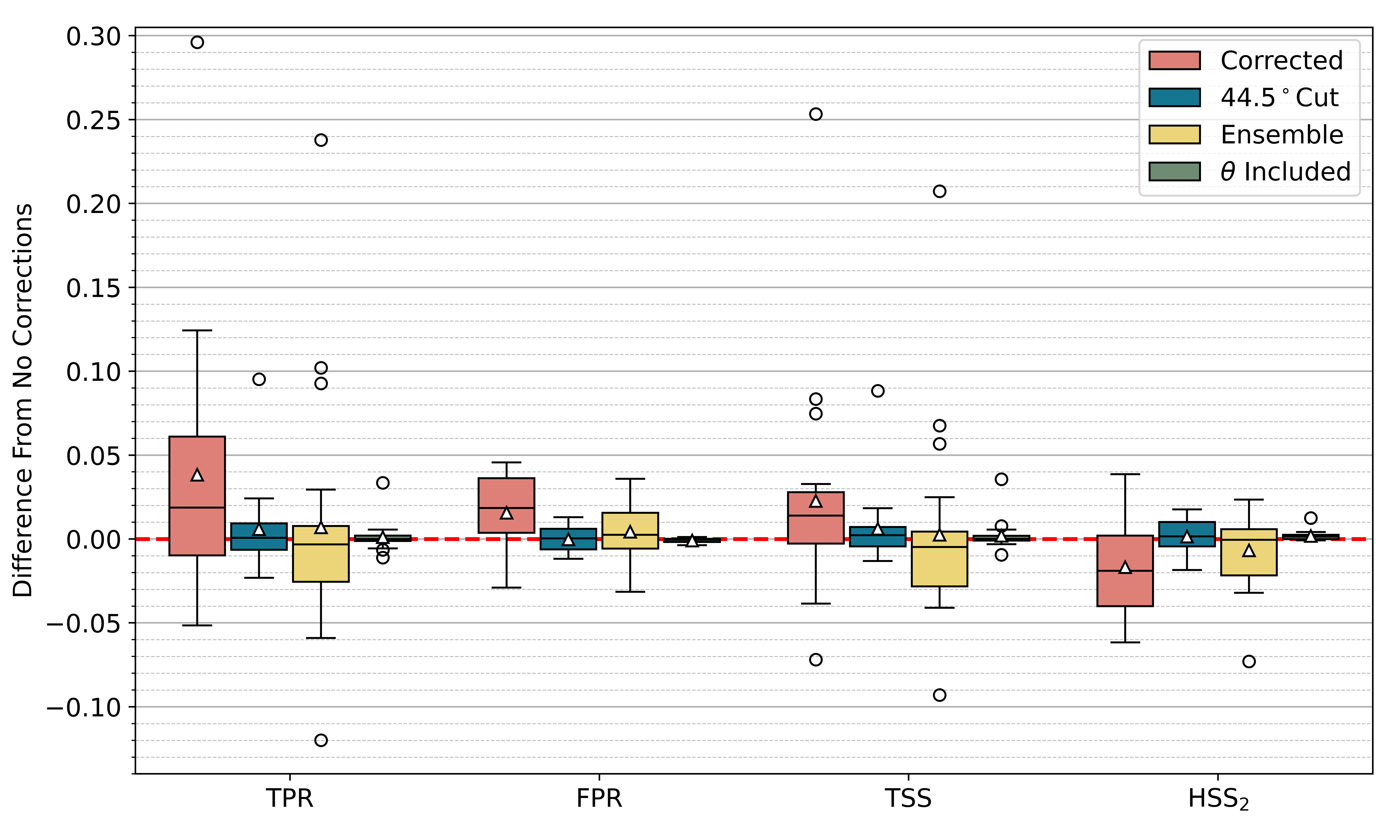}
    \caption{The true positive rate (TPR), false positive rate (FPR), true skill statistic (TSS), and Heidke skill score (HSS$_2$) differences from the uncorrected approach for every training/testing partition combination from SWAN-SF. Positive values indicate an increase in the score from the baseline of no corrections. Negative values indicate a decrease from the baseline. Results are color-coded to indicate whether the classifier was trained/tested with the corrected data (coral), the ``44.5° Cut" methodology (blue), the ensemble methodology (yellow), or the uncorrected data with the mean heliocentric angle (green). The red dashed line emphasizes no difference from the uncorrected classifier. See the caption in Figure \ref{fig:scoresScatter} for the meaning of the triangles and circles in the box plots.}
    \label{fig:ensembleHA}
\end{figure}

Aside from directly manipulating the data, we address projection effects in two additional ways. First, we employ an ensemble-based forecaster dependent on the mean $\theta$ of ARs. This involves training three SVM classifiers on three different $\theta$ ranges: [0$^\circ$ - 23.33$^\circ$], (23.33$^\circ$ - 46.66$^\circ$], and (46.66$^\circ$ - 70.00$^\circ$]. Within each $\theta$ range, we are minimizing the long-term $\theta$-dependent projection effect trends within a particular training data set, which, in a way, removes the impacts of these effects without altering the data itself. However, we are also removing the number of flaring samples used to train an individual model, which is a huge downside. When making forecasts, the appropriate model is then selected based on the location of the observed AR. For each classifier, we utilize the same feature selection and hyperparameter tuning methodology mentioned in Section \ref{sec:method}. 

We also test an approach that retrains the uncorrected classifiers with the mean $\theta$ as a feature. This ensures our model considers the location of the AR when making a forecast. These classifiers were trained using the same selected features (with the addition of the mean $\theta$) and hyperparameters originally chosen for the uncorrected data sets. 

The results for both tests are shown in Figure \ref{fig:ensembleHA}. Here we display a boxplot noting the TPR, FPR, TSS, and HSS$_2$ deviations from no corrections for the corrected data (maroon), the ``44.5° Cut" methodology (salmon pink), the ensemble methodology (teal), and the uncorrected data plus mean $\theta$ (dark blue). These differences are calculated across the same training/testing partition pairs. Once again, our alternate methods for considering projection effects have little impact on performance. The ensemble methodology shows a slight increase in TSS scores with a complimentary decrease in HSS$_2$. The changes are on the order of only a few percent. Finally, when including the mean $\theta$ as an additional feature to the uncorrected classifier, performance is nearly unchanged across all tested metrics. 
\label{sec:otherPotential} \label{sec:summary}
\section{Summary}
Vector magnetograms suffer from projection effects as active regions migrate toward the limb. This can cause problems in solar flare forecasting, as active regions may appear stronger or weaker than they truly are due to artificial long-term $\theta$-dependent trends present within the data. In this work, we perform several correction techniques to remedy this problem and study their impact on flare forecasting. From our analysis, we have made the following conclusions:
\begin{enumerate}
    \item Correcting for projection effects using a data-centered methodology increases both the true positive and false positive prediction rates. However, these changes are minimal.
    \item Beyond heliocentric angles of 44.5$^\circ$, ``good" and ``bad" changes to our forecasts after corrections are applied tend to level out. If we would like to maximize performance improvements, a combination of the corrected and uncorrected classifiers should be used. For data with mean heliocentric angles below 44.5$^\circ$, the classifier trained on uncorrected data should be used. For heliocentric angles above 44.5$^\circ$, the corrected classifier should be used.
    \item Classifier-based correction methodologies showed no clear improvements over data-centered corrections. 
    \item A more sophisticated correction methodology is likely needed to see greater improvements in our results.
    \item These findings could indicate significant limitations in using magnetogram data for flare forecasting. This further emphasizes the need for new data sources and prediction techniques to continue the development of the space weather forecasting community.
\end{enumerate}

Finally, it could be intriguing to approach this problem from a modeling perspective in the future. Simulating an active region and feeding it through the Solar Dynamics Observatory Helioseismic and Magnetic Imager pipeline \citep[similar to][]{Kitiashvili_Couvidat_Lagg_2015} would allow us to better constrain the projection effects for each magnetic field parameter. We could force the simulated active regions into a particular shape, size, and strength, allowing us to generate a more robust selection of ``benchmark" active regions.

\section{Acknowledgements}
We want to thank the anonymous reviewer for their valuable comments, which greatly improved the clarity of the text. This work was supported by NASA FINESST grant 80NSSC23K1639, NASA LWS grant 80NSSC22K0272, and NSF FDSS grant 1936361. Viacheslav Sadykov acknowledges the NASA COFFIES DSC grant NNH18ZDA001N-DRIVE.

\bibliography{sample631}{}
\bibliographystyle{aasjournal}



\end{document}